\documentclass[a4paper,UKenglish,cleveref, autoref]{lipics-v2019}

\usepackage{algorithm2e}
\usepackage{bm}
\usepackage{mathtools}

\newcommand{\set}[1]{\left\{#1\right\}}
\newcommand{\NN}{{\mathbb{N}}}
\newcommand{\ZZ}{{\mathbb{Z}}}
\newcommand{\RR}{{\mathbb{R}}}
\newcommand{\CC}{{\mathbb{C}}}
\newcommand{\QQ}{{\mathbb{Q}}}
\newcommand{\TT}{{\mathbb{T}}}

\newcommand{\aff}{{\mathrm{aff}}}
\newcommand{\conv}{{\mathrm{conv}}}

\newcommand{\bgamma}{{\boldsymbol{\gamma}}}
\newcommand{\bmu}{{\boldsymbol{\mu}}}
\newcommand{\btheta}{{\boldsymbol{\theta}}}
\newcommand{\bpsi}{{\boldsymbol{\psi}}}

\newcommand{\bv}{\bm{v}}
\newcommand{\bx}{\bm{x}}
\newcommand{\bz}{\bm{z}}

\newcommand{\bb}{\bm{b}}
\newcommand{\bc}{\bm{c}}
\newcommand{\bp}{\bm{p}}

\newcommand{\Ploop}{\mathsf{P}}
\newcommand{\Qloop}{\mathsf{Q}}

\title{Termination of Linear Loops over the Integers} 

\author{Mehran Hosseini}{Department of Computer Science, University of Oxford, UK }{mehran.hosseini@cs.ox.ac.uk}{}{Mehran Hosseini was supported by ERC grant AVS-ISS (648701).}

\author{Jo\"el Ouaknine}{Max Planck Institute for Software Systems, Germany \and Department of Computer Science, University of Oxford, UK}{joel@mpi-sws.org}{}{Jo\"el Ouaknine was supported by ERC grant AVS-ISS (648701) and by DFG grant 389792660 as part of TRR~248 (see \url{https://perspicuous-computing.science}).}

\author{James Worrell}{Department of Computer Science, University of Oxford, UK}{jbw@cs.ox.ac.uk}{}{James Worrell was supported by EPSRC Fellowship EP/N008197/1.}

\authorrunning{M. Hosseini, J. Ouaknine, and J. Worrell}

\Copyright{Mehran Hosseini, Jo\"el Ouaknine, and James Worrell}



\nolinenumbers 

\EventEditors{Christel Baier, Ioannis Chatzigiannakis, Paola Flocchini, and Stefano Leonardi}
\EventNoEds{4}
\EventLongTitle{46th International Colloquium on Automata, Languages, and Programming (ICALP 2019)}
\EventShortTitle{ICALP 2019}
\EventAcronym{ICALP}
\EventYear{2019}
\EventDate{July 9--12, 2019}
\EventLocation{Patras, Greece}
\EventLogo{eatcs}
\SeriesVolume{132}
\ArticleNo{114}

\begin{document}

\maketitle

\begin{abstract}
  We consider the problem of deciding termination of single-path while
  loops with integer variables, affine updates, and affine guard
  conditions. The question is whether such a loop terminates on all
  integer initial values.  This problem is known to be decidable for
  the subclass of loops whose update matrices are diagonalisable, but
  the general case has remained open since being conjectured decidable
  by Tiwari in 2004. In this paper we show decidability of determining
  termination for arbitrary update matrices, confirming Tiwari's
  conjecture.  For the class of loops considered in this paper, the
  question of deciding termination on a single initial value is a
  longstanding open problem in number theory.  The key to our decision
  procedure is in showing how to circumvent the difficulties inherent
  in deciding termination on a single initial value.
\end{abstract}

\section{Introduction}
\label{sec:intro}
Termination is a central problem in program verification. In this
paper we study termination of \emph{single-path affine loops}, i.e.,
programs of the form
\begin{gather*}
  \mathit{while} \; (B \bx > \bb) \; \mathit{do} \;
  \bx:= A \bx + \bc \, ,
\end{gather*}
where \(A, B, \bb, \bc\) are matrices of appropriated dimensions.
Such loop programs are often referred to as \emph{linear loops}.
Here the loop body has a single control path that performs a
simultaneous affine update of the program variables.  Analysis of
loops of this form, including acceleration and termination, is an
important part of analysing more complex programs (see,
e.g.,~\cite{Boigelot03,JeannetSS14,KincaidBCR19}).

For a set \(S\subseteq \RR^d\), we say that the above loop
\emph{terminates on \(S\)} if it terminates on all initial values in
\(S\).  Despite the simplicity of single-path affine loops, the
question of deciding termination has proven challenging (and
termination becomes undecidable if the update function in the loop
body is allowed to be piecewise linear or if the loop body consists of
a nondeterministic choice between two different linear
updates~\cite{Ben-AmramGM12}).  Tiwari~\cite{T04} showed that
termination of single-path affine loops is decidable over
\(\RR^d\). Subsequently, Braverman~\cite{B06}, using a more refined
analysis of the loop components, showed that termination is decidable
over \(\QQ^d\) and noted that termination on \(\mathbb{Z}^d\) can be
reduced to termination on \(\QQ^d\) in the homogeneous case, i.e.,
when \(\bb, \bc\) are both all-zero vectors.  More recently, Ouaknine,
Sousa-Pinto, and Worrell~\cite{OPW15} have proven that termination
over \(\ZZ^d\) is decidable in the non-homogeneous case under the
assumption that the update
matrix \(A\) is a diagonalisable integer matrix. Decidability of
termination for non-homogeneous affine loops over \(\mathbb{Z}^d\) was
conjectured by Tiwari~\cite[Conjecture 1]{T04}, but has remained open
until now.

In this paper we give a procedure for deciding termination of the
general class of single-path affine loops over the integers, i.e., we
generalise the result of~\cite{OPW15} by lifting the assumption of
diagonalisability. Note that for this class of programs, the question
of termination on a single initial value in \(\mathbb{Z}^d\) (as
opposed to termination over all of \(\mathbb{Z}^d\)) is equivalent to
the \emph{Positivity Problem} for linear recurrence sequences, i.e.,
the problem of whether all terms in a given integer linear recurrence
sequence are positive. Decidability of the Positivity Problem is a
longstanding open problem (going back at least as far as the
1970s~\cite{RS94,Soi76}), and results in~\cite{OW14a} suggest that a
solution to the problem will require significant breakthroughs in
number theory. However, in considering termination over
\(\mathbb{Z}^d\) we show that one can benefit from the freedom to
choose the initial values of the loop variables. In the present paper
we exploit this freedom in order to circumvent the need to solve
``hard instances'' of the Positivity Problem when deciding termination
of affine loops. In particular, we avoid the use of sophisticated
Diophantine-approximation techniques, such as the \(S\)-units theorem,
that were employed in~\cite{OW14a}. By eschewing such tools we lose
all hope of obtaining an effective characterisation of the set of
non-terminating points. (Compare with the approach in~\cite{OPW15},
which yielded an effective characterisation of the set of all
eventually non-terminating points in the diagonalisable case.)
Nevertheless our methods manage to solve the decision problem in the
general case.

Among the tools we use are a circle of closely related classical
results on the geometry of numbers, including Khinchine's flatness
theorem, Kronecker's theorem on simultaneous Diophantine
approximation, and the result of Khachiyan and Porkolab that it is
decidable whether a convex semi-algebraic set contains an integer
point.  In tandem with these, from algebraic number theory, we use a
result of Masser that allows to compute all algebraic relations among
the eigenvalues of the update matrix of a given loop.  Using this last
result, we define a semi-algebraic subset of ``non-termination
candidates'' such that the loop is non-terminating if and only if this
set contains an integer point.

In this paper we focus on the foundational problem of providing
complete methods to solve termination.  Much effort has been devoted
to scalable and pragmatic methods to prove termination for classes of
programs that subsume affine loops.  In particular, techniques to
prove termination via synthesis of linear ranking
functions~\cite{BG13,BG14,BMS05,CFM15,CS01,PR04-1,PR04-2} and their
extension, multiphase linear ranking functions~\cite{BG17,BDG18}, have
been developed.  Many of these techniques have been implemented in
software verification tools, such as Microsoft's
\textsc{Terminator}~\cite{CPR06}. Although these methods are capable
of handling non-deterministic affine loops, they can only guarantee
termination whenever ranking functions of a certain form exist.

\section{Background}
\subsection{Exponential Polynomials}
\label{sec:exp-poly}
Let \(\lambda_1,\ldots,\lambda_m \in \mathbb{C}\) be distinct complex
numbers and \(e_1,\ldots,e_m\) positive integers.  Then the family of
\emph{exponential-polynomial} functions
\(p_{i,j} : \mathbb{N}\rightarrow \mathbb{C}\), for
\(j\in\{1,\ldots,m\}\) and \(i\in\{0,\ldots,e_j-1\}\), given by
\(p_{i,j}(n) = \binom{n}{i}\lambda_j^n\) is linearly independent over
\(\mathbb{C}\).  Moreover if \(p:\mathbb{N}\rightarrow\mathbb{C}\) is
a \(\mathbb{C}\)-linear combination of the \(p_{i,j}\), then \(p\) is
identically zero iff \(p(n)=0\) for \(e_1+\cdots+e_m\) consecutive
values \(n\in \mathbb{N}\).  Both of the above facts can be proved
using generalised Vandermonde determinants~\cite[Proposition
2.11]{TUCS05}.

\subsection{Convexity}
The \emph{affine hull} of \(S\subseteq \mathbb{R}^d\) is the smallest
affine set that contains \(S\), where an affine set is the translation
of a vector subspace of \(\mathbb{R}^d\).  The affine hull of \(S\)
can be characterised as follows:
\begin{equation*}
  \aff(S) := \left\{ \sum_{i=1}^k \alpha_i \boldsymbol x_i
    \mid k>0,\boldsymbol x_i \in S,\alpha_i \in \mathbb{R},
    \sum_{i=1}^k \alpha_i =1 \right \} \, .
\end{equation*}
The \emph{convex hull} of \(S\subseteq \mathbb{R}^d\) is the smallest
convex set that contains \(S\).  The convex hull of \(S\) can be
characterised as follows:
\begin{equation*}
  \conv(S) := \left\{ \sum_{i=1}^k \alpha_i \boldsymbol x_i
    \mid k>0,\boldsymbol x_i \in S,\alpha_i \in \mathbb{R}_{\geq 0},
    \sum_{i=1}^k \alpha_i =1 \right \} \, .
\end{equation*}
Clearly \(\conv(S)\subseteq \aff(S)\).  The \emph{relative interior}
of a convex set \(S\subseteq \mathbb{R}^d\) is its interior with
respect to the restriction of the Euclidean topology to \(\aff(S)\).
We have the following easy proposition, characterising the relative
interior.

\begin{proposition}
  Let
  \(S = \{\boldsymbol a_1,\ldots,\boldsymbol a_n\} \subseteq
  \mathbb{R}^d\).  If \(\boldsymbol u\) lies in the relative interior
  of \(\conv(S)\) then there exist \(\alpha_1,\ldots,\alpha_n > 0\)
  such that \(u=\sum_{i=1}^n \alpha_i \boldsymbol a_i\) and
  \(\sum_{i=1}^n \alpha_i = 1\).
\label{prop:relative-int}
\end{proposition}
\begin{proof}
  Since \(\boldsymbol u\) lies in the relative interior of
  \(\conv(S)\), for \(\varepsilon > 0\) sufficiently small we
  have that
  \begin{equation*} (1+n\varepsilon)u - \sum_{i=1}^n \varepsilon
    \boldsymbol a_i \in \conv(S) \, .
  \end{equation*}
  For such an \(\varepsilon\) there exist
  \(\beta_1,\ldots,\beta_n \geq 0\) such that
  \((1+n\varepsilon)u - \sum_{i=1}^n \varepsilon \boldsymbol a_i =
  \sum_{i=1}^n \beta_i \boldsymbol a_i\) and
  \(\sum_{i=1}^n \beta_i = 1\).  But then
  \(\boldsymbol u = \sum_{i=1}^n \frac{\beta_i +
    \varepsilon}{1+n\varepsilon} \boldsymbol a_i\).  Defining
  \(\alpha_i:= \frac{\beta_i + \varepsilon}{1+n\varepsilon}\) for
  \(i\in \{1,\ldots,n\}\), the proposition is proved.
\end{proof}

A \emph{lattice of rank \(r\)} in \(\RR^d\) is a set 
\begin{equation*}
  \Lambda :=\{ z_1 \boldsymbol v_1 + \cdots + z_r \boldsymbol v_r
  \colon z_1,\ldots,z_r \in \mathbb{Z}\} \, ,
\end{equation*}
where \(\boldsymbol v_1,\ldots,\boldsymbol v_r\) are linearly
independent vectors in \(\RR^d\).  Given a convex set
\(C\subseteq \RR^d\), define the \emph{width} of \(C\) along a vector
\(\boldsymbol u \in \RR^d\) to be
\begin{equation*} \sup\{ \boldsymbol u^\top (\boldsymbol x -
  \boldsymbol y) : \boldsymbol x,\boldsymbol y \in C \} \, .
\end{equation*}
Furthermore the \emph{lattice width} of \(C\) is the infimum over all
non-zero vectors \(\boldsymbol u \in \Lambda\) of the width of \(C\)
along \(\boldsymbol u\).

The following result (see~\cite{flatness99, Khinchin48}) captures the
intuition that a convex set that contains no lattice point in its
interior must be ``thin'' in some direction.
\begin{theorem}[Flatness Theorem]
  Given a full-rank lattice \(\Lambda\) in \(\mathbb{R}^d\) there
  exists \(W\) such that any convex set \(C\subseteq \RR^d\) that has
  non-empty interior and lattice width at least \(W\) contains a
  lattice point in its interior.
\label{thm:flatness}
\end{theorem}

Recall that \(C\subseteq \RR^d\) is said to be \emph{semi-algebraic}
if it is definable by a boolean combination of polynomial constraints
\(p(x_1,\ldots,x_d) > 0\), where \(p \in \ZZ[x_1,\ldots,x_d]\).

\begin{theorem}[Khachiyan and Porkolab~\cite{KP97}]
  \label{thm:KP}
  It is decidable whether a given convex semi-algebraic set
  \(C\subseteq \RR^d\) contains an integer point, that is, whether
  \(C \cap \ZZ^d \neq \emptyset\).
\end{theorem}

\subsection{Groups of Multiplicative Relations}
\label{subsec:grp_of_mul_rel}
In this subsection we will introduce some concepts concerning groups
of multiplicative relations among algebraic numbers.

Let \(\TT = \set{z \in \CC : |z|=1 }\). We define the
\(s\)-dimensional torus to be \(\TT^s\), considered as a group under
component-wise multiplication.  Given a tuple of algebraic numbers
\(\bgamma = (\gamma_1, \cdots, \gamma_s) \in \TT^s\), the orbit
\(\set{\bgamma^n : n\in\NN}\) is a subset of \(\TT^s\).  In the
following we characterise the topological closure of the orbit as an
algebraic subset of \(\TT^s\).

The \emph{group of multiplicative relations} of \(\bgamma\in \TT^s\)
is defined as the following additive subgroup of \(\ZZ^s\):
\begin{equation*}
L(\bgamma)=\set{\bv\in\ZZ^s:\bgamma^{\bv}=1},
\end{equation*}
where \(\bgamma^{\bv}\) is defined to be
\(\gamma_1^{v_1}\cdots\gamma_s^{v_s}\) for \(\bv\in\ZZ^s\), that is,
exponentiation acts coordinate-wise.  Since \(L(\bgamma)\) is a
subgroup of \(\ZZ^s\), it is a free Abelian group and hence has a
finite basis.  The following powerful theorem of Masser \cite{Mas88}
gives bounds on the magnitude of the components of such a basis.

\begin{theorem}[Masser]
  \label{thm:Mas}
  The free Abelian group \(L(\bgamma)\) has a basis
  \(\bv_1,\ldots,\bv_l\in\ZZ^s\) for which
  \begin{equation*}
    \max_{1\leq i \leq l, 1 \leq j \leq s}|v_{i,j}|\leq(D\log H)^{O(s^2)},
  \end{equation*}
  where \(H\) and \(D\) bound respectively the heights and degrees of
  all the \(\gamma_i\).
\end{theorem}

Membership of a tuple \(\bv \in \ZZ^s\) in \(L(\bgamma)\) can be
computed in polynomial time, using exponentiation by squaring
method. In combination with \autoref{thm:Mas}, it follows that we can
compute a basis for \(L(\bgamma)\) in polynomial space by brute-force
search.

Corresponding to \(L(\bgamma)\), we consider the following
multiplicative subgroup of \(\TT^s\):
\begin{equation*}
  T(\bgamma)=\set{\bmu \in \TT^s : \forall\bv \in L(\bgamma),
    \bmu^{\bv}=1}.
\end{equation*}
If \(\mathcal{B}\) is a basis of \(L(\bgamma)\), we can equivalently
characterise \(T(\bgamma)\) as
\(\set{\bmu \in \TT^s : \forall \bv \in \mathcal{B},
  \bmu^{\bv}=1}\). Crucially, this finitary characterisation allows us
to represent \(T(\bgamma)\) as an algebraic set in \(\mathbb{T}^s\).

We will use the following classical lemma of Kronecker on simultaneous
Diophantine approximation to show that the orbit
\(\set{\bgamma^n : n \in \NN}\) is a dense subset of \(T(\bgamma)\).

\begin{lemma}
  \label{lem:Kron}
  Let \(\btheta, \bpsi \in \RR^s\). Suppose that for all
  \(\bv \in \ZZ^s\), if \(\bv^\top \btheta \in \ZZ\) then also
  \(\bv^\top \bpsi \in \ZZ\), i.e., all integer relations among the
  coordinates of \(\btheta\) also hold among those of \(\bpsi\)
  (modulo \(\ZZ\)). Then, for each \(\varepsilon > 0\), there exist
  \(\bp \in \ZZ^s\) and a non-negative integer \(n\) such that
  \begin{equation*}
    \Vert n\btheta-\bp-\bpsi \Vert_\infty \leq \varepsilon.
  \end{equation*}
\end{lemma}

We now arrive at the main result of the section:
\begin{theorem}
  Let \(\bgamma \in \TT^s\). Then the orbit
  \(\set{\bgamma^k : k \in \NN}\) is a dense subset of \(T(\bgamma)\).
  \label{thm:dense}
\end{theorem}
\begin{proof}
  Let \(\btheta \in \RR^s\) be such that
  \(\bgamma = e^{2\pi i \btheta}\) (with exponentiation operating
  coordinate-wise). Notice that \(\bgamma^{\bv} = 1\) if and only if
  \(\bv^\top \btheta \in \ZZ\). If \(\bmu \in T(\bgamma)\), we can
  likewise define \(\bpsi \in \RR^s\) to be such that
  \(\bmu = e^{2\pi i\bpsi}\) . Then the premises of Kronecker's lemma
  apply to \(\btheta\) and \(\bpsi\). Thus, given \(\varepsilon > 0\),
  there exist a non-negative integer \(k\) and \(\bp \in \ZZ^s\) such
  that \(\Vert k\btheta - \bp - \bpsi \Vert_\infty \leq
  \varepsilon\). Whence
  \begin{equation*}
    \Vert \bgamma^k - \bmu \Vert_\infty = \Vert e^{2\pi i (k\btheta-\bp)}
    - e^{2\pi i \bpsi} \Vert_\infty \leq \Vert 2\pi(k\btheta - \bp -
    \bpsi) \Vert_\infty \leq 2\pi\varepsilon.
  \end{equation*}
\end{proof}

\section{Termination Analysis via Spectral Theory}
\label{sec:spectral}
The general form of a single-path affine loop in dimension \(d\) is as
follows:
\begin{gather*}
  \mathit{while} \; (g_1(\boldsymbol x) > 0 \wedge \ldots \wedge
  g_m(\boldsymbol x) > 0) \; \mathit{do} \; \boldsymbol
  x:=f(\boldsymbol x) \, ,
\end{gather*}
where \(g_1,\ldots,g_m : \mathbb{R}^d \rightarrow \mathbb{R}\) and
\(f : \mathbb{R}^d \rightarrow \mathbb{R}^d\) are affine functions.
We assume that \(f\) and \(g_1,\ldots,g_m\) have integer coefficients,
that is, \(f(\boldsymbol x)=A\boldsymbol x + \boldsymbol a\) for
\(A \in \mathbb{Z}^{d\times d}\) and
\(\boldsymbol a \in \mathbb{Z}^d\), and
\(g_i(\boldsymbol x) = \boldsymbol b_i^\top \boldsymbol x + c_i\) for
\(\boldsymbol b_i \in \mathbb{Z}^d\), \(c_i \in \mathbb{Z}\) and
\(i=1,\ldots,m\).

Note that 
\begin{gather}
  \begin{pmatrix} f(\boldsymbol x) \\ 1 \end{pmatrix} =
  \begin{pmatrix} A& \boldsymbol a\\ 0 & 1 \end{pmatrix}
  \begin{pmatrix} \boldsymbol x \\ 1 \end{pmatrix} \text{ and }
  g_i(\boldsymbol x) = (\boldsymbol b_i^\top \; c_i)
  \begin{pmatrix} \boldsymbol x \\ 1 \end{pmatrix} \, .
  \label{eq:update}
\end{gather}
for all \(\boldsymbol x \in \mathbb{R}^d\).  We say that \(f\) is
\emph{non-degenerate} if no quotient of two distinct eigenvalues of
the update matrix
\(\begin{psmallmatrix} A& \boldsymbol a\\ 0 & 1 \end{psmallmatrix}\)
is a root of unity.

\begin{proposition}
  The termination problem for single-path affine loops on integers is
  reducible to the special case of the problem for non-degenerate
  update functions.
\end{proposition}
\begin{proof}
  Consider a single-path affine loop, as described above, whose update
  matrix has distinct eigenvalues \(\lambda_1,\ldots,\lambda_s\).  Let
  \(L\) be the least common multiple of the orders of the roots of
  unity appearing among the quotients \(\frac{\lambda_i}{\lambda_j}\)
  for \(i\neq j\).  It is known that
  \(L = 2^{O(d\sqrt{\log d})}\)~\cite[subsection~1.1.9]{EPSW03}.  The
  update matrix corresponding to the affine map
  \(f^L = \underbrace{f \circ \cdots \circ f}_L\) has eigenvalues
  \(\lambda_1^L,\ldots,\lambda_s^L\) and hence is non-degenerate.
  Moreover the original loop terminates if and only if the following
  loop terminates:
  \begin{gather*}
    \mathit{while} \; \bigwedge_{i=0}^{L-1} \left(g_1(f^i(\boldsymbol
      x)) > 0 \wedge \ldots \wedge g_m(f^i(\boldsymbol x)) > 0\right)
    \; \mathit{do} \; \boldsymbol x:=f^L(\boldsymbol x) \, ,
  \end{gather*}
  This concludes the proof.
\end{proof}

In the rest of this section and in the next section we focus on the
case of a loop
\begin{gather}
  \Ploop \,\colon\, \mathit{while} \; (g(\boldsymbol x)>0)
  \;\mathit{do} \; \boldsymbol x \leftarrow f(\boldsymbol x)
\label{eq:the-loop}
\end{gather}
with a single guard function
\(g(\boldsymbol x)=\boldsymbol b^\top \boldsymbol x + c\) and with
non-degenerate update function
\(f(\boldsymbol x)=A\boldsymbol x+\boldsymbol a\), with both maps
having integer coefficients.  We show that a spectral analysis of the
matrix underlying the loop update function suffices to classify almost
all initial values of the loop as either terminating or eventually
non-terminating.  Towards the end of the section we isolate a class of
so-called \emph{critical} initial values that are not amenable to this
analysis.  We show how to deal with such points in
\autoref{sec:critical}.

With respect to the loop \(\Ploop\) we say that
\(\boldsymbol{x} \in \RR^d\) is \emph{terminating} if there exists
\(n\) such that \(g(f^n(\boldsymbol x)) \leq 0\).  We say that
\(\boldsymbol x\) is \emph{eventually non-terminating} if the sequence
\(\langle g(f^n(\boldsymbol x)) : n\in\mathbb{N}\rangle\) is
\emph{ultimately positive}, i.e., there exists \(N\) such that for all
\(n\geq N\), \(g(f^n(\boldsymbol x)) > 0\).  Clearly there exists
\(\boldsymbol z \in \mathbb{Z}^d\) that is non-terminating if and only
if there exists \(\boldsymbol z \in \mathbb{Z}^d\) that is eventually
non-terminating.  Thus we can regard the problem of deciding
termination on \(\mathbb{Z}^d\) as that of searching for an eventually
non-terminating point.

Let \(\lambda_1,\ldots,\lambda_s\) be the non-zero eigenvalues of
\(\begin{psmallmatrix} A& \boldsymbol a\\ 0 & 1 \end{psmallmatrix}\)
and let \(k_{\mathrm{max}}\) be the maximum multiplicity over all
these eigenvalues.

Define a linear preorder on
\(I:=\{0,\ldots,k_{\mathrm{max}}-1\} \times \{1,\ldots,s\}\) by
\((i_1,j_1) \preccurlyeq (i_2,j_2)\) if either
(i)~\(|\lambda_{j_1}| <|\lambda_{j_2}|\) or
(ii)~\(|\lambda_{j_1}|=|\lambda_{j_2}|\) and \(i_1 \leq i_2\).  Write
\((i_1,j_1) \prec (i_2,j_2)\) if \((i_1,j_1) \preccurlyeq (i_2,j_2)\)
and \((i_2,j_2) \not\preccurlyeq (i_1,j_1)\).  Then we have
\begin{equation*}
  (i_1,j_1) \prec (i_2,j_2) \text{ iff } \lim_{n\rightarrow \infty}
  \frac{ \binom{n}{i_1}|\lambda_{j_1}|^n}{
    \binom{n}{i_2}|\lambda_{j_2}|^n } = 0 \, ,
\end{equation*}
that is, the preorder \(\preccurlyeq\) characterises the asymptotic
order of growth in absolute value of the terms
\(\binom{n}{i}\lambda^n_j\) for \((i,j) \in I\).  This preorder
moreover induces an equivalence relation \(\approx\) on \(I\) where
\((i_1,j_1) \approx (i_2,j_2)\) iff
\((i_1,j_1) \preccurlyeq (i_2,j_2)\) and
\((i_2,j_2) \preccurlyeq (i_1,j_1)\).

The following closed-form expression for \(g(f^n(\boldsymbol x))\)
will be the focus of the subsequent development.
\begin{proposition}
  There is a set of affine functions
  \(h_{i,j} : \mathbb{R}^d \rightarrow \mathbb{C}\) such that for all
  \(\boldsymbol x \in \mathbb{R}^d\) and all \(n \geq d\) we have
  \(\boldsymbol g(f^n(\boldsymbol{x})) = \sum_{(i,j)\in I}
  \binom{n}{i} \lambda_j^n \, h_{i,j}(\boldsymbol{x})\).
\label{prop:ind}
\end{proposition}
\begin{proof}
  Using the Jordan-Chevalley decomposition, we can write
  \(\begin{psmallmatrix} A & \boldsymbol a\\ 0 & 1 \end{psmallmatrix}
  =P^{-1}DP+N\), where \(D\) is diagonal, \(N\) is nilpotent, \(P\) is
  invertible, \(P^{-1}DP\) and \(N\) commute, and all matrices have
  algebraic coefficients.  Moreover we can write
  \(D=\lambda_1 D_1 + \cdots + \lambda_s D_s\) for appropriate
  idempotent diagonal matrices \(D_1,\ldots,D_s\).  Then for all
  \(n\in\mathbb{N}\) with \(n\geq d\) we have
  \begin{align}
    g(f^n(\boldsymbol x))
    & = (\boldsymbol{b}^\top \; c)
      \begin{pmatrix} A
        & \boldsymbol a\\ 0 & 1 \end{pmatrix}^n 
                              \begin{pmatrix} \boldsymbol{x} \\ 1 
                              \end{pmatrix}  \nonumber \\
    & = (\boldsymbol{b}^\top \; c) (P^{-1}DP+N)^n
      \begin{pmatrix}  
        \boldsymbol{x} \\ 1
      \end{pmatrix}  \nonumber \\
    & = (\boldsymbol{b}^\top \; c) \sum_{i=0}^n \binom{n}{i} P^{-1}
      D^{n-i} PN^i
      \begin{pmatrix}   \boldsymbol{x} \\ 1  \end{pmatrix}   \nonumber \\
    & = 
      (\boldsymbol{b}^\top \; c) \sum_{i=0}^{d} \binom{n}{i} P^{-1}
      (\lambda_1^{n-i}D_1 + \cdots + \lambda_s^{n-i}D_s )PN^i 
      \begin{pmatrix}  
        \boldsymbol{x} \\ 1
      \end{pmatrix} & \text{(since \(N^{d+1}=0\))} \nonumber \\
    & = \sum_{j=1}^s \lambda_j^n \sum_{i=0}^{d} \binom{n}{i}
      \underbrace{\lambda_j^{-i} (\boldsymbol{b}^\top \; c)
      P^{-1}D_jPN^i \begin{pmatrix} \boldsymbol{x} \\ 1
      \end{pmatrix} }_{h_{i,j}(\boldsymbol x)} \label{eq:def-h}\\
    & = \sum_{j=1}^s \sum_{i=0}^{d} \binom{n}{i} \lambda_j^n
      h_{i,j}(\boldsymbol{x}) \, , \nonumber
\end{align} 
where for \((i,j)\in I\) the affine function \(h_{i,j}\) is defined in
Line~\eqref{eq:def-h}.  Clearly each function \(h_{i,j}\) is a
complex-valued affine function on \(\mathbb{R}^d\) with algebraic
coefficients.
\end{proof}

Define \(\gamma_i = \frac{\lambda_i}{|\lambda_i|}\) for
\(i=1,\ldots,s\), that is, we obtain the \(\gamma_i\) by normalising
the eigenvalues to have length \(1\).  Recall from
\autoref{subsec:grp_of_mul_rel} the definition of the group
\(L(\bgamma)\) of multiplicative relations that hold among
\(\gamma_{1},\ldots,\gamma_{s}\), \emph{viz.},
\begin{equation*}
  L(\bgamma) = \{ (n_1,\ldots,n_s) \in \mathbb{Z}^s :
  \gamma_1^{n_1} \cdots \gamma_s^{n_s} = 1 \} \, .
\end{equation*}
Recall also that we have \(T(\bgamma) \subseteq \mathbb{T}^s\), given by 
\begin{equation*} T(\bgamma) = \{ (\mu_1,\ldots,\mu_s) \in
  \mathbb{T}^s : \mu_1^{n_1} \cdots \mu_s^{n_s} = 1 \text{ for all }
  (n_1,\ldots,n_s) \in L(\bgamma) \} \, .
\end{equation*}

Given an \(\approx\)-equivalence class \(L \subseteq I\), note that
for all \((i_1,j_1),(i_2,j_2) \in L\) we have \(i_1=i_2\) and
\(|\lambda_{j_1}|=|\lambda_{j_2}|\).  Thus \(L\) is determines a
common multiplicity, which we denote \(i_L\), and a set of eigenvalues
that all have the same absolute value, which we denote \(\rho_L\).

Given an \(\approx\)-equivalence class \(L\), define
\(\Phi_L : \mathbb{R}^d \times T(\bgamma) \rightarrow \mathbb{R}\)
by\footnote{That the function \(\Phi_L\) is real-valued follows from
  the fact that if eigenvalues \(\lambda_{j_1}\) and \(\lambda_{j_2}\)
  are complex conjugates then \(\gamma_{j_1}\) and \(\gamma_{j_2}\)
  are also complex conjugates, as are \(h_{i,j_1}(\boldsymbol z)\) and
  \(h_{i,j_2}(\boldsymbol z)\) (see the proof of \autoref{prop:ind}).}
\begin{gather}
  \Phi_L(\boldsymbol x,\boldsymbol \mu) = \sum_{(i,j)\in L}
  h_{i,j}(\boldsymbol x) \mu_j \, .
\label{def:Phi}
\end{gather}
From the above definition of \(\Phi_L\) we have
\begin{gather}
  \sum_{(i,j) \in L} \binom{n}{i}\lambda_j^n h_{i,j}(\boldsymbol x) =
  \binom{n}{i_L}\rho_L^n \Phi_L(\boldsymbol x,\boldsymbol \gamma^n) \,
  .
  \label{eq:formula}
\end{gather}
for all \(\boldsymbol x \in \mathbb{R}^d\) and all
\(n\in \mathbb{N}\).

We say that an equivalence class \(E\) of \(I\) is \emph{dominant} for
\(\boldsymbol x \in \mathbb{R}^d\) if \(E\) is the equivalence class
of the maximal indices \((i,j)\) for which \(h_{i,j}(\boldsymbol x)\)
is non-zero.  Equivalently, \(E\) is dominant for \(\boldsymbol x\) if
\(E\) is the maximal equivalence class such that
\(\Phi_E(\boldsymbol x,\cdot)\) is not identically zero on
\(T(\bgamma)\).  (The equivalence of these two characterisations
follows from the linear independence of the functions
\(\binom{n}{i}\lambda_j^n\) for \((i,j) \in E\).)

The following proposition shows how information about termination of
the loop \(\Ploop\) on an initial value \(\boldsymbol x\in\RR^d\) can
be derived from properties of \(\Phi_E(\boldsymbol x,\cdot)\).

\begin{proposition}\label{prop: easy iv's}
  Consider the loop \(\Ploop\) in~\eqref{eq:the-loop}.  Let
  \(\boldsymbol x \in \mathbb{R}^d\) and let \(E\) be an
  \(\approx\)-equivalence class that is dominant for
  \(\boldsymbol x\).  Then
  \begin{enumerate}
  \item\label{itm: easy non-term iv's} If
    \(\displaystyle\inf_{\boldsymbol{\mu} \in T(\bgamma)}
    \Phi_E(\boldsymbol x,\boldsymbol \mu) > 0\) then
    \(\boldsymbol{x}\) is eventually non-terminating for~\(\Ploop\).
  \item\label{itm: easy term iv's} If
    \(\displaystyle\inf_{\boldsymbol{\mu} \in T(\bgamma)}
    \Phi_E(\boldsymbol x,\boldsymbol \mu) < 0\) then
    \(\boldsymbol{x}\) is terminating for~\(\Ploop\).
  \end{enumerate}
\end{proposition}
\begin{proof}
  By \autoref{prop:ind} and Equation~\eqref{eq:formula} we have that
  for all \(n\geq d\),
  \begin{eqnarray}
    g(f^n(\boldsymbol x))
    &=& \sum_{(i,j) \in I} \binom{n}{i}\lambda_j^n h_{i,j}(\boldsymbol{x}) \notag \\
    &=& \binom{n}{i_E} \rho_E^n \Phi_E(\boldsymbol x,\boldsymbol \gamma^n) +
        \sum_{(i,j) \in I\setminus E} \binom{n}{i}\lambda_j^n h_{i,j}(\boldsymbol{x}) \, .
  \label{eqn:dom}
  \end{eqnarray}
  Moreover by the dominance of \(E\) we have that
  \begin{gather} \lim_{n\rightarrow \infty}
    \frac{\binom{n}{i}|\lambda_j|^n}{\binom{n}{i_E}\rho_E^n} = 0
    \label{eq:dominate}
  \end{gather}
  for all \((i,j) \in I \setminus E\) such that
  \(h_{i,j}(\boldsymbol x)\neq 0\).

  We first prove \autoref{itm: easy non-term iv's}.  By assumption, in
  this case there exists \(\varepsilon>0\) such that
  \(\Phi_E(\boldsymbol x,\boldsymbol \mu) \geq \varepsilon\) for all
  \(\boldsymbol \mu \in T(\bgamma)\).  Together with
  Equation~\eqref{eq:dominate}, this shows that the asymptotically
  dominant term in Equation~\eqref{eqn:dom} has positive sign.  It
  follows that \(g(f^n(\boldsymbol x))\) is positive for \(n\)
  sufficiently large and hence \(\boldsymbol x\) is eventually
  non-terminating.

  We turn now to \autoref{itm: easy term iv's}.  By assumption there
  exists \(\varepsilon>0\) and an open subset \(U\) of \(T(\bgamma)\)
  such that \(\Phi_E(\boldsymbol x,\boldsymbol \mu) < -\varepsilon\)
  for all \(\boldsymbol \mu \in U\).  Moreover by density of
  \(\{ \boldsymbol \gamma^n : n \in \mathbb{N}\}\) in \(T(\bgamma)\)
  there exist infinitely many \(n\) such that
  \(\boldsymbol \gamma^n \in U\).  Exactly as in Case 1 we can now use
  the dominance of \(E\) to conclude that \(g(f^n(\boldsymbol x))<0\)
  for sufficiently large \(n\) such that
  \(\boldsymbol \gamma^n \in U\) and hence \(\boldsymbol x\) is
  terminating.
\end{proof}

Given \(\boldsymbol z \in \mathbb{Z}^d\), since \(T(\bgamma)\) is an
algebraic subset of \(\TT^s\), the number
\(\displaystyle\inf_{\boldsymbol{\mu} \in T(\bgamma)}
\Phi_E(\boldsymbol z,\boldsymbol \mu)\) is algebraic and its sign can
be decided.  Note however that \autoref{prop: easy iv's} does not
completely resolve the question of termination with respect to guard
\(g\) from a given initial value \(\boldsymbol z\).  Indeed, let us
define \(\boldsymbol z \in \mathbb{R}^d\) to be \emph{critical} if
\(\displaystyle\inf_{\boldsymbol{\mu} \in E} \Phi_E(\boldsymbol
z,\boldsymbol \mu) = 0\), where \(E\) is the dominant equivalence
class for \(\boldsymbol z\).  Then neither clause in the above
proposition suffices to resolve termination of the loop \(\Ploop\)
in~\eqref{eq:the-loop} on such a \(\boldsymbol z\).  Indeed the
question of whether such a point is eventually non-terminating is
equivalent to the \emph{Ultimate Positivity Problem} for linear
recurrence sequences: a longstanding and notoriously difficult open
problem in number theory, only known to be decidable up to order 4
\cite{ACHOW18, OW14a}. Fortunately in the setting of deciding loop
termination we can sidestep such difficult questions.  The following
section is devoted to handling critical points.  The idea is to show
that if there is a critical initial value then there is another
initial value that is eventually non-terminating and moreover whose
eventual non-termination can be established by \autoref{prop: easy
iv's}.

\section{Analysis of Critical Points}
\label{sec:critical}
In this section we continue to analyse termination of the loop
\(\Ploop\), as given in~\eqref{eq:the-loop} in the previous section,
and refer to the notation established therein.

\subsection{Transition Invariance of Critical Points}
Intuitively critical points are those for which it is difficult to
determine eventual non-termination.  One should therefore expect that
if \(\boldsymbol x \in \RR^d\) is critical then \(f(\boldsymbol x)\)
should also be critical.  This, and more, follows from the following
proposition.

\begin{proposition}
  Let \(\boldsymbol x \in \mathbb{R}^d\) and let \(E\subseteq I\) be
  an equivalence class that is dominant for \(\boldsymbol x\).  Then
  \(E\) is also dominant for \(f(\boldsymbol x)\), and for all
  \(\boldsymbol \mu \in T(\bgamma)\) we have
  \(\Phi_E(f(\boldsymbol x),\boldsymbol \mu) = \rho_E \,
  \Phi_E(\boldsymbol x,\boldsymbol \gamma \boldsymbol \mu)\), where
  the product \(\boldsymbol \gamma \boldsymbol \mu\) is defined
  pointwise.
  \label{prop:shift}
\end{proposition}
\begin{proof}
  By definition we have
  \(\Phi_E(\boldsymbol x,\boldsymbol \mu) = \sum_{(i,j)\in E}
  h_{i,j}(\boldsymbol x)\mu_j\), where the \(h_{i,j}\) satisfy
\begin{gather}
  (\boldsymbol b^\top \; c) \begin{pmatrix}A& \boldsymbol a \\ 0 &
    1 \end{pmatrix}^n
  \begin{pmatrix} \boldsymbol x \\ 1 \end{pmatrix} = \sum_{(i,j) \in
    I} h_{i,j}(\boldsymbol x) \binom{n}{i}\lambda_j^n \,
  \label{eq:first}
\end{gather}
for all \(n\geq d\).  Likewise we have
\(\Phi_E(f(\boldsymbol x),\boldsymbol \mu) = \sum_{(i,j)\in E}
\widetilde{h}_{i,j}(\boldsymbol x)\mu_j\), where the
\(\widetilde{h}_{i,j}\) satisfy
\begin{gather} (\boldsymbol b^\top \; c) \begin{pmatrix}A& \boldsymbol
    a \\ 0 & 1 \end{pmatrix}^{n+1}
  \begin{pmatrix} \boldsymbol x \\ 1 \end{pmatrix} = \sum_{(i,j) \in
    I} \widetilde{h}_{i,j}(\boldsymbol x) \binom{n}{i}\lambda_j^n \, .
\label{eq:second}
\end{gather}
Combining Equations~\eqref{eq:first} and~\eqref{eq:second} we have the
for all \(n\geq d\),
\begin{eqnarray*}
  \sum_{(i,j) \in I} \widetilde{h}_{i,j}(\boldsymbol x) \binom{n}{i} \lambda_j^n
  &=& 
        \sum_{(i,j) \in I} h_{i,j}(\boldsymbol x) \binom{n+1}{i}\lambda_j^{n+1} \\
  &=& \sum_{(i,j) \in I} h_{i,j}(\boldsymbol x)
      \left[\binom{n}{i}+\binom{n}{i-1}\right]\lambda_j \lambda_j^{n} \, .
\end{eqnarray*}

Now the collection of functions \(n\mapsto \binom{n}{i}\lambda_j^n\)
for \((i,j) \in I\) is linearly independent (see
\autoref{sec:exp-poly}).  Equating the coefficients of the functions
\(\binom{n}{i}\lambda_j^n\) for \((i,j) \in E\) in the above equation
we have
\(\widetilde{h}_{i,j} = \lambda_j h_{i,j} = \rho_E \gamma_j h_{i,j}\)
for all \((i,j) \in E\); likewise we have that \(E\) is dominant for
\(f(\boldsymbol x)\).  The proposition follows.
\end{proof}

The next lemma shows that the existence of a critical point entails
the existence of an eventually non-terminating point.

\begin{lemma}
  If \(\boldsymbol z \in \mathbb{R}^d\) is critical then for all
  \(n \geq 2d+1\), all points in the relative interior of
  \(\conv(\{ f^d(\boldsymbol z),f^{d+1}(\boldsymbol
  z),\ldots,f^{n}(\boldsymbol z)\})\) are eventually non-terminating.
  \label{lem:critical}
\end{lemma}
\begin{proof}
  Let \(E\) be the \(\approx\)-equivalence class that is dominant for
  \(\boldsymbol z\).  Fix \(\bmu \in T(\bgamma)\).  We claim that
  there exists \(n \geq d\) such that \(\Phi_E (f^n(\bz),\bmu) >
  0\). If this were not the case then by \autoref{prop:shift} for all
  \(n \geq d\) we would have
  \(\Phi_E (f^n(\bz),\bmu) = \rho_E^n \, \Phi_E
  (\bz,\bgamma^n\bmu)=0\).  But by \autoref{thm:dense}, the set
  \(\{ \bgamma^n\bmu : n \geq d\}\) is dense in \(T(\bgamma)\) and
  hence we would have that \(\Phi_E (\bz,\cdot)\) is identically \(0\)
  on \(T(\bgamma)\), contradicting the dominance of \(E\).  This
  establishes the claim.

  In fact we can sharpen the above claim to state that for some
  \(n\in \{d,d+1,\ldots,2d+1\}\) we have
  \(\Phi_E (f^n(\bz),\bmu) > 0\).  Indeed for all \(n\geq d\) we have
  \begin{equation*}
    \Phi_E(f^n(\boldsymbol z),\boldsymbol \mu) = \rho_E^n
    \Phi_E(\boldsymbol z,\boldsymbol \gamma^n \boldsymbol \mu) =
    \sum_{(i,j) \in E} h_{i,j}(\boldsymbol z) \rho_E^n \gamma_j^n
    \mu_j \, .
  \end{equation*}
  Thus the sequence
  \(\langle \Phi_E(f^n(\boldsymbol z),\bmu) : n\geq d \rangle\) can be
  written as a sum of exponentials with at most \(d+1\) terms.  Since
  this sequence is not identically zero, it has a non-zero entry for
  some \(n\in \{d,d+1,\ldots,2d+1\}\) (cf.\ \autoref{sec:exp-poly}).
  Since \(\bmu\) was arbitrary, we have that for all
  \(\bmu \in T(\bgamma)\) there exists \(n\in \{d,d+1,\ldots,2d+1\}\)
  with \(\Phi_E (f^n(\bz),\bmu) > 0\).

  By \autoref{prop:relative-int}, for all \(n \geq 2d+1\) and all
  points \(\boldsymbol x\) lying in the relative interior of
  \(\conv(\{ f^d(\boldsymbol z),f^{d+1}(\boldsymbol
  z),\ldots,f^{n}(\boldsymbol z)\})\), there exist
  \(\alpha_d,\ldots,\alpha_n > 0\) such that
  \(\sum_{i=d}^n \alpha_i =1\) and
  \(\boldsymbol x = \sum_{i=d}^n \alpha_i f^i(\boldsymbol z)\).  Since
  \(\Phi_E\) is an affine map in its first variable, it follows that
  \(\Phi_E(\boldsymbol x,\cdot) = \sum_{i=d}^n \alpha_i \Phi_E
  (f^i(\boldsymbol z),\cdot)\) is strictly positive on \(T(\bgamma)\).
  Hence \(\boldsymbol x\) is eventually non-terminating by
  \autoref{prop: easy iv's}.
\end{proof}

\subsection{Finding Integer Non-Terminating Points from Critical Points}
\autoref{lem:critical} shows how to derive the existence of
non-terminating points from the existence of a critical point.  In
this subsection we refine this analysis to derive the existence of
\emph{integer} non-terminating points.  In particular, fixing an
initial value \(\boldsymbol z_* \in \mathbb{Z}^d\), we show that for
\(n\) sufficiently large, the set
\begin{equation*}
  \conv(\{f^d(\boldsymbol z_*), f^{d+1}(\boldsymbol
  z_*),\ldots,f^{n}(\boldsymbol z_*)\})
\end{equation*}
contains an integer point in its relative interior.

Define \(V:= \aff(\{ f^n (\boldsymbol z_*) : n \geq d\})\) and let the
vector subspace \(V_0 \subseteq \RR^d\) be the unique translate of
\(V\) containing the origin.  Write \(d_0\) for the dimension of
\(V_0\) (equivalently the dimension of \(V\)).

\begin{proposition}
  For all non-zero integer vectors \(\boldsymbol v \in V_0\) the set
  \(\{ |\boldsymbol v^\top f^{n}(\boldsymbol z_*)| : n \geq d \}\) is
  unbounded.
  \label{prop:unbounded}
\end{proposition}
\begin{proof}
  Consider the sequence
  \(x_n:=\boldsymbol v^\top f^n(\boldsymbol z_*) = v^\top
  \begin{psmallmatrix} A&\boldsymbol a\\ 0& 1 \end{psmallmatrix}^n
  \begin{psmallmatrix} \boldsymbol z_* \\ 1 \end{psmallmatrix}\).  If
  this sequence were constant then \(\boldsymbol v\) would be
  orthogonal to \(V_0\), contradicting the fact that \(\boldsymbol v\)
  is a non-zero vector in \(V_0\).  Since the sequence is
  non-constant, integer-valued, and satisfies a non-degenerate linear
  recurrence of order at most \(d+1\) (see,
  e.g.,~\cite[subsection~1.1.12]{EPSW03}), by the Skolem-Mahler-Lech
  Theorem we have that
  \(\{ |\boldsymbol v^\top f^{n}(\boldsymbol z_*)| : n \geq d \}\) is
  unbounded (see the discussion of growth of linear recurrence
  in~\cite[section~2.2]{EPSW03}).\footnote{The above argument actually
    establishes that \(\langle x_n : n \in\mathbb{N}\rangle\) diverges
    to infinity in absolute value.  We briefly sketch a more
    elementary proof of mere unboundedness.  If the sequence
    \(\langle x_n : n \in\mathbb{N}\rangle\) were bounded then by van
    der Waerden's Theorem, for all \(m\) it would contain a constant
    subsequence of the form \(x_\ell,x_{\ell+p},\ldots,x_{\ell+mp}\)
    for some \(\ell,p \geq 1\).  In particular, if \(m=d\) then since
    every infinite subsequence \(y_n:=x_{\ell+pn}\) satisfies a linear
    recurrence of order at most \(d+1\),
    \(\langle x_n : n \in \NN \rangle\) would have an infinite
    constant subsequence \(\langle x_{\ell+pn} : n\in \NN \rangle\).
    If \(p=1\) then \(\langle x_n : n \in \NN \rangle\) is constant
    and if \(p>1\) then by~\cite[Lemma 9.11]{SS78}
    \(\langle x_n : n \in \NN \rangle\) is degenerate.}
\end{proof}

\begin{proposition}
  There exists \(M\) such that for all \(n\geq M\) the set
  \begin{equation*}
    \conv(\{f^d(\boldsymbol z_*),f^{d+1}(\boldsymbol
    z_*),\ldots,f^n(\boldsymbol z_*)\})
  \end{equation*}
  contains an integer point in its relative interior.
  \label{prop:rel-int}
\end{proposition}
\begin{proof}
  Since \(V_0\) is spanned by integer vectors,
  \(\Lambda:=V_0\cap\ZZ^d\) is a lattice of rank \(d_0\) in \(\RR^d\).
  Define \(C:=\conv(\{ f^n(\boldsymbol z_*) : n\geq d\}) \subseteq V\)
  and \(C_0:=C-f^d(\boldsymbol z_*) \subseteq V_0\).

  Let \(\theta:\RR^d \rightarrow \RR^{d_0}\) be a linear map that
  takes \(V_0\) bijectively onto \(\RR^{d_0}\) and whose kernel is the
  orthogonal complement of \(V_0\). Then \(\theta(\Lambda)\) is a
  lattice in \(\RR^{d_0}\) of full rank.  We claim that the lattice
  width of \(\theta(C_0)\) with respect to \(\theta(\Lambda)\) is
  infinite.  Indeed for any non-zero vector
  \(\boldsymbol v \in \theta(\Lambda)\) we have
  \begin{gather}
    \boldsymbol v^\top (\theta(f^n(\boldsymbol z_*)) -
    \theta(f^d(\boldsymbol z_*))) = (\theta^*\boldsymbol v)^\top
    (f^n(\boldsymbol z_*)-f^d(\boldsymbol z_*)) \, ,
    \label{eq:UB}
  \end{gather} where \(\theta^*:\RR^{d_0}\rightarrow \RR^d\) is the
  adjoint map of \(\theta\).  But \(\theta^*\boldsymbol v\) is a
  non-zero rational vector in \(V_0\) and hence
  \autoref{prop:unbounded} entails that the absolute value
  of~\eqref{eq:UB} is unbounded as \(n\) runs over \(\mathbb{N}\).
  This proves the claim.

  Since \(\theta(C_0)\) is a full-dimensional convex subset of
  \(\mathbb{R}^{d_0}\), by \autoref{thm:flatness} we have that
  \(\theta(C_0)\) contains a point of \(\theta(\Lambda)\) in its
  relative interior and hence \(C_0\) contains a point of \(\Lambda\)
  (necessarily an integer point) in its relative interior.  We
  conclude that \(C\) also contains an integer point in its relative
  interior.
\end{proof}

We summarise sections~\ref{sec:spectral} and~\ref{sec:critical} with a
theorem characterising when a loop with a single guard is terminating.

\begin{theorem}
  The loop \(\Ploop\), given in~\eqref{eq:the-loop}, is
  non-terminating on \(\ZZ^d\) if and only if there exists
  \(\boldsymbol z \in \ZZ^d\) and an \(\approx\)-equivalence class
  \(E\) such that (i)~\(E\) is dominating for \(\boldsymbol z\) and
  (ii)~\(\displaystyle\inf_{\boldsymbol \mu \in T(\bgamma)}
  \Phi_E(\boldsymbol z,\boldsymbol \mu) \geq 0\).
\label{thm:one-guard}
\end{theorem}
\begin{proof}
  If no such \(\boldsymbol z\) exists then the loop is terminating by
  \autoref{prop: easy iv's}.(\ref{itm: easy term iv's}).  Conversely,
  if such a \(\boldsymbol z\) exists then by \autoref{lem:critical}
  and \autoref{prop:rel-int} there exists
  \(\boldsymbol z' \in \mathbb{Z}^d\) such that
  \(\displaystyle\inf_{\boldsymbol \mu \in T(\bgamma)}
  \Phi_E(\boldsymbol z',\boldsymbol \mu) > 0\) (and with \(E\) still
  dominating for \(\boldsymbol z'\).)  Such a point is eventually
  non-terminating by \autoref{prop: easy iv's}.(\ref{itm: easy
    non-term iv's}).
\end{proof}

We postpone the question of the effectiveness of the above
characterisation until we handle loops with multiple guards, in
\autoref{sec:multiple}.

\section{Multiple Guards}
\label{sec:multiple}
Now we are ready to present our decision procedure for a general
affine loop program
\begin{gather}
  \Qloop \, : \, \mathit{while} \; (g_1(\boldsymbol x) > 0 \wedge
  \ldots \wedge g_m(\boldsymbol x) > 0) \; \mathit{do} \; \boldsymbol
  x:=f(\boldsymbol x) \, ,
  \label{eq:multi-loop}
\end{gather}
with multiple guards.  Associated to the loop \(\Qloop\) we consider
\(m\) single-guard loops with a common update function:
\begin{gather*}
  \Qloop_i \, : \, \mathit{while} \; (g_i(\boldsymbol x) > 0) \;
  \mathit{do} \; \boldsymbol x:=f(\boldsymbol x) \, ,
\end{gather*}
for \(i=1,\ldots,m\).  Clearly \(\Qloop\) is non-terminating if and
only if there exists \(\boldsymbol z \in \ZZ^d\) such that each loop
\(\Qloop_i\) is non-terminating on \(\boldsymbol z\).  As we now
explain, we can decide the existence of such a point following the
proof of \autoref{thm:one-guard}.

Let \(\lambda_1,\ldots,\lambda_s\) be the distinct non-zero
eigenvalues of the matrix corresponding to the update function \(f\)
in the loop \(\Qloop\).  As before, write
\(\gamma_j = {\lambda_j}/{|\lambda_j|}\) for \(j=1,\ldots,s\).  For
\(i=1,\ldots,m\), denote by
\(\Phi^{(i)}_E : \mathbb{R}^d\times T(\bgamma) \rightarrow \RR\) the
function associated to loop \(\Qloop_i\) and \(\approx\)-equivalence
class \(E\) as defined by~\eqref{def:Phi}.  Given
\(\approx\)-equivalence classes \(E_1,\ldots,E_m\), we define
\(W_{E_1,\ldots,E_m}\subseteq\RR^d\) to be the set of
\(\bgamma \in \RR^d\) such that the following hold for
\(i=1,\ldots,m\):
\begin{itemize}
\item[•] \(E_i\) is dominant for \(\bx\) in loop \(\Qloop_i\), that
  is, \(\Phi_{E_i}^{(i)}(\boldsymbol x,\cdot) \not\equiv 0\) and
  \(\Phi_{E}^{(i)}(\boldsymbol x,\cdot) \equiv 0\) for all
  \(E_i \prec E\).
\item[•]
  \(\displaystyle \inf_{\boldsymbol \mu \in T(\bgamma)}
  \Phi^{(i)}_{E_i}(\boldsymbol x,\boldsymbol \mu) \geq 0\).
\end{itemize}

\begin{proposition}
  Loop \(\Qloop\) is non-terminating if and only if there exist
  \(\approx\)-equivalence classes \(E_1,\ldots,E_m\) such that
  \(W_{E_1,\ldots,E_m}\) contains an integer point.
\label{prop:char}
\end{proposition}
\begin{proof}
  Suppose that \(\Qloop\) fails to terminate on
  \(\boldsymbol z \in \ZZ^d\).  Then each loop \(\Qloop_i\) also fails
  to terminate on \(\boldsymbol z \in \ZZ^d\).  Thus if \(E_i\) is the
  dominant equivalence class for \(\boldsymbol z\) in program
  \(\Qloop_i\), for \(i=1,\ldots,m\), applying \autoref{prop: easy
    iv's}.(\ref{itm: easy term iv's}) we get that
  \(\boldsymbol z \in W_{E_1,\ldots,E_m}\).

  Conversely, suppose \(\boldsymbol z \in W_{E_1,\ldots,E_m}\) for
  some \(\approx\)-equivalence classes \(E_1,\ldots,E_m\).  Then, by
  \autoref{lem:critical} and \autoref{prop:rel-int}, there is an
  integer point
  \(\boldsymbol z' \in \conv(\{ f^n(\boldsymbol z) : n \geq d\})\)
  such that
  \(\displaystyle \inf_{\boldsymbol \mu \in T(\bgamma)}
  \Phi^{(i)}_{E_i} (\boldsymbol z',\boldsymbol \mu) > 0\) for
  \(i=1,\ldots,m\).  By \autoref{prop: easy iv's}.(\ref{itm: easy
    non-term iv's}), each loop \(\Qloop_i\) fails to terminate on
  \(\boldsymbol z'\) and hence also \(\Qloop\) is non-terminating on
  \(\boldsymbol z'\).
\end{proof}

~\autoref{prop:char} leads to the following procedure for deciding
termination of a given affine loop \(\Qloop\), as shown
in~\eqref{eq:multi-loop}.
\begin{enumerate}
\item Compute the eigenvalues of the matrix corresponding to the loop
  update function, as given in~\eqref{eq:update}.
\item Compute the dominance preorder \(\preccurlyeq\) among
  eigenvalues.
\item Compute a basis of the group of multiplicative relations
  \(L(\bgamma)\).
\item Return ``non-terminating'' if some set \(W_{E_1,\ldots,E_m}\)
  contains an integer point and otherwise return ``terminating''.
\end{enumerate}

In terms of effectiveness, Steps 1 and 2 can be accomplished via
standard symbolic computations with algebraic numbers.  (We refer
to~\cite{OPW15} for a detailed treatment in a very similar setting.)
By \autoref{thm:Mas}, computing a basis of \(L(\bgamma)\) reduces to
checking a finite collection of multiplicative relations among
algebraic numbers.  Given a basis of \(L(\bgamma)\) we can directly
obtain representations of each set \(W_{E_1,\ldots,E_m}\) as
semi-algebraic subsets of \(\RR^d\).  Finally, since
\(W_{E_1,\ldots,E_m}\) is convex, we can decide the existence of an
integer point in each set \(W_{E_1,\ldots,E_m}\) using
\autoref{thm:KP}.

We have thus established the main result of the paper:
\begin{theorem}
  There is a procedure to decide termination of single-path affine
  loops (of the form specified in~\eqref{eq:multi-loop}) over the
  integers.
\end{theorem}

\bibliography{lipics-v2019-sample-article}

\begin{thebibliography}{10}

\bibitem{ACHOW18}
Shaull Almagor, Brynmor Chapman, Mehran Hosseini, Jo{\"{e}}l Ouaknine, and
  James Worrell.
\newblock Effective divergence analysis for linear recurrence sequences.
\newblock In {\em 29th International Conference on Concurrency Theory, {CONCUR}
  2018, September 4-7, 2018, Beijing, China}, pages 42:1--42:15, 2018.

\bibitem{flatness99}
Wojciech Banaszczyk, Alexander~E Litvak, Alain Pajor, and Stanislaw~J Szarek.
\newblock The flatness theorem for nonsymmetric convex bodies via the local
  theory of banach spaces.
\newblock {\em Mathematics of operations research}, 24(3):728--750, 1999.

\bibitem{BDG18}
Amir~M. Ben{-}Amram, Jes{\'{u}}s Dom{\'{e}}nech, and Samir Genaim.
\newblock Multiphase-linear ranking functions and their relation to recurrent
  sets.
\newblock {\em CoRR}, abs/1811.07340, 2018.

\bibitem{BG13}
Amir~M. Ben{-}Amram and Samir Genaim.
\newblock On the linear ranking problem for integer linear-constraint loops.
\newblock In {\em The 40th Annual {ACM} {SIGPLAN-SIGACT} Symposium on
  Principles of Programming Languages, {POPL} '13, Rome, Italy - January 23 -
  25, 2013}, pages 51--62, 2013.

\bibitem{BG14}
Amir~M. Ben{-}Amram and Samir Genaim.
\newblock Ranking functions for linear-constraint loops.
\newblock {\em J. {ACM}}, 61(4):26:1--26:55, 2014.

\bibitem{BG17}
Amir~M. Ben{-}Amram and Samir Genaim.
\newblock On multiphase-linear ranking functions.
\newblock In {\em Computer Aided Verification - 29th International Conference,
  {CAV} 2017, Heidelberg, Germany, July 24-28, 2017, Proceedings, Part {II}},
  pages 601--620, 2017.

\bibitem{Ben-AmramGM12}
Amir~M. Ben{-}Amram, Samir Genaim, and Abu~Naser Masud.
\newblock On the termination of integer loops.
\newblock {\em {ACM} Trans. Program. Lang. Syst.}, 34(4):16:1--16:24, 2012.

\bibitem{Boigelot03}
Bernard Boigelot.
\newblock On iterating linear transformations over recognizable sets of
  integers.
\newblock {\em Theor. Comput. Sci.}, 309(1-3):413--468, 2003.

\bibitem{BMS05}
Aaron~R. Bradley, Zohar Manna, and Henny~B. Sipma.
\newblock Termination analysis of integer linear loops.
\newblock In {\em {CONCUR} 2005 - Concurrency Theory, 16th International
  Conference, {CONCUR} 2005, San Francisco, CA, USA, August 23-26, 2005,
  Proceedings}, pages 488--502, 2005.

\bibitem{B06}
Mark Braverman.
\newblock Termination of integer linear programs.
\newblock In {\em Computer Aided Verification, 18th International Conference,
  {CAV} 2006, Seattle, WA, USA, August 17-20, 2006, Proceedings}, pages
  372--385, 2006.

\bibitem{CFM15}
Hong~Yi Chen, Shaked Flur, and Supratik Mukhopadhyay.
\newblock Termination proofs for linear simple loops.
\newblock {\em {STTT}}, 17(1):47--57, 2015.

\bibitem{CS01}
Michael Col{\'{o}}n and Henny Sipma.
\newblock Synthesis of linear ranking functions.
\newblock In {\em Tools and Algorithms for the Construction and Analysis of
  Systems, 7th International Conference, {TACAS} 2001 Held as Part of the Joint
  European Conferences on Theory and Practice of Software, {ETAPS} 2001 Genova,
  Italy, April 2-6, 2001, Proceedings}, pages 67--81, 2001.

\bibitem{CPR06}
Byron Cook, Andreas Podelski, and Andrey Rybalchenko.
\newblock Termination proofs for systems code.
\newblock In {\em Proceedings of the {ACM} {SIGPLAN} 2006 Conference on
  Programming Language Design and Implementation, Ottawa, Ontario, Canada, June
  11-14, 2006}, pages 415--426, 2006.

\bibitem{EPSW03}
Graham Everest, Alfred~J. van~der Poorten, Igor~E. Shparlinski, and Thomas
  Ward.
\newblock {\em Recurrence Sequences}, volume 104 of {\em Mathematical surveys
  and monographs}.
\newblock American Mathematical Society, 2003.

\bibitem{TUCS05}
V.~Halava, T.~Harju, M.~Hirvensalo, and J.~Karhum\"aki.
\newblock Skolem's problem -- on the border between decidability and
  undecidability.
\newblock Technical Report 683, Turku Centre for Computer Science, 2005.

\bibitem{JeannetSS14}
Bertrand Jeannet, Peter Schrammel, and Sriram Sankaranarayanan.
\newblock Abstract acceleration of general linear loops.
\newblock In {\em The 41st Annual {ACM} {SIGPLAN-SIGACT} Symposium on
  Principles of Programming Languages, {POPL} '14, San Diego, CA, USA, January
  20-21, 2014}, pages 529--540. {ACM}, 2014.

\bibitem{KP97}
Leonid Khachiyan and Lorant Porkolab.
\newblock Computing integral points in convex semi-algebraic sets.
\newblock In {\em 38th Annual Symposium on Foundations of Computer Science,
  {FOCS} '97, Miami Beach, Florida, USA, October 19-22, 1997}, pages 162--171,
  1997.

\bibitem{Khinchin48}
Aleksandr~Yakovlevich Khinchin.
\newblock Dirichlet's principle in the theory of diophantine approximations.
\newblock {\em Uspekhi Matematicheskikh Nauk}, 3(3):3--28, 1948.

\bibitem{KincaidBCR19}
Zachary Kincaid, Jason Breck, John Cyphert, and Thomas~W. Reps.
\newblock Closed forms for numerical loops.
\newblock {\em {PACMPL}}, 3({POPL}):55:1--55:29, 2019.

\bibitem{Mas88}
David~W Masser.
\newblock Linear relations on algebraic groups.
\newblock {\em New Advances in Transcendence Theory}, pages 248--262, 1988.

\bibitem{OPW15}
Jo{\"{e}}l Ouaknine, Jo{\~{a}}o~Sousa Pinto, and James Worrell.
\newblock On termination of integer linear loops.
\newblock In {\em Proceedings of the Twenty-Sixth Annual {ACM-SIAM} Symposium
  on Discrete Algorithms, {SODA} 2015, San Diego, CA, USA, January 4-6, 2015},
  pages 957--969, 2015.

\bibitem{OW14a}
Jo{\"{e}}l Ouaknine and James Worrell.
\newblock Positivity problems for low-order linear recurrence sequences.
\newblock In {\em Proceedings of the Twenty-Fifth Annual {ACM-SIAM} Symposium
  on Discrete Algorithms, {SODA} 2014, Portland, Oregon, USA, January 5-7,
  2014}, pages 366--379, 2014.

\bibitem{PR04-1}
Andreas Podelski and Andrey Rybalchenko.
\newblock A complete method for the synthesis of linear ranking functions.
\newblock In {\em Verification, Model Checking, and Abstract Interpretation,
  5th International Conference, {VMCAI} 2004, Venice, Italy, January 11-13,
  2004, Proceedings}, pages 239--251, 2004.

\bibitem{PR04-2}
Andreas Podelski and Andrey Rybalchenko.
\newblock Transition invariants.
\newblock In {\em 19th {IEEE} Symposium on Logic in Computer Science {(LICS}
  2004), 14-17 July 2004, Turku, Finland, Proceedings}, pages 32--41, 2004.

\bibitem{RS94}
G.~Rozenberg and A.~Salomaa.
\newblock {\em Cornerstones of Undecidability}.
\newblock Prentice Hall, 1994.

\bibitem{SS78}
Arto Salomaa and Matti Soittola.
\newblock {\em Automata-Theoretic Aspects of Formal Power Series}.
\newblock Texts and Monographs in Computer Science. Springer, 1978.

\bibitem{Soi76}
M.~Soittola.
\newblock On {D0L} synthesis problem.
\newblock In A.~Lindenmayer and G.~Rozenberg, editors, {\em Automata,
  Languages, Development}. North-Holland, 1976.

\bibitem{T04}
Ashish Tiwari.
\newblock Termination of linear programs.
\newblock In {\em Computer Aided Verification, 16th International Conference,
  {CAV} 2004, Boston, MA, USA, July 13-17, 2004, Proceedings}, pages 70--82,
  2004.

\end{thebibliography}

\appendix

\end{document}